\documentclass{article}
\usepackage{arxiv}
\usepackage[utf8]{inputenc} 
\usepackage[T1]{fontenc}    
\usepackage{hyperref}       
\usepackage{url}            
\usepackage{booktabs}       
\usepackage{graphicx}
\usepackage{doi}

\usepackage{algorithmic}
\usepackage[ruled,linesnumbered]{algorithm2e}
\usepackage{latexsym}
\usepackage{amssymb}
\usepackage{amsmath}
\usepackage{amsthm}
\usepackage{booktabs}
\usepackage{subcaption}
\usepackage{graphicx}
\usepackage{multirow}
\usepackage{color}
\usepackage{tikz}
\usepackage{comment}

\date{}

\title{Physics-Guided Variational Model for Unsupervised Sound Source Tracking}

\author{Luan Vinícius Fiorio~\thanks{Corresponding author.} \\
	Department of Electrical Engineering\\
	Eindhoven University of Technology\\
    De Groene Loper 19\\
	5600 MB Eindhoven, The Netherlands \\
	\texttt{l.v.fiorio@tue.nl} \\
    \And
	Ivana Nikoloska \\
	Department of Electrical Engineering\\
	Eindhoven University of Technology\\
    De Groene Loper 19\\
	5600 MB Eindhoven, The Netherlands \\
	\texttt{i.nikoloska@tue.nl} \\
    \And
    Bruno Defraene \\
	NXP Semiconductors\\
	Interleuvenlaan 80\\
	3001 Leuven, Belgium\\
	\texttt{bruno.defraene@nxp.com} \\    
    \And
    Alex Young \\
	NXP Semiconductors\\
	High Tech Campus 60\\
	5656 AE Eindhoven, The Netherlands\\
	\texttt{alex.young@nxp.com} \\
	\And
    Johan David \\
	NXP Semiconductors\\
	Interleuvenlaan 80\\
	3001 Leuven, Belgium\\
	\texttt{j.david@nxp.com} \\
    \And
	Ronald M. Aarts\\
	Department of Electrical Engineering\\
	Eindhoven University of Technology\\
    De Groene Loper, 19\\
	5600 MB Eindhoven, The Netherlands \\     
    \texttt{R.M.Aarts@tue.nl} \\
}

\renewcommand{\shorttitle}{\textit{arXiv} Template}

\hypersetup{
pdftitle={A template for the arxiv style},
pdfsubject={q-bio.NC, q-bio.QM},
pdfauthor={David S.~Hippocampus, Elias D.~Striatum},
pdfkeywords={First keyword, Second keyword, More},
}

\begin{document}
\maketitle

\begin{abstract}
Sound source tracking is commonly performed using classical array-processing algorithms, while machine-learning approaches typically rely on precise source position labels that are expensive or impractical to obtain. This paper introduces a physics-guided variational model capable of fully unsupervised single-source sound source tracking. The method combines a variational encoder with a physics-based decoder that injects geometric constraints into the latent space through analytically derived pairwise time-delay likelihoods. Without requiring ground-truth labels, the model learns to estimate source directions directly from microphone array signals. Experiments on real-world data demonstrate that the proposed approach outperforms traditional baselines and achieves accuracy and computational complexity comparable to state-of-the-art supervised models. We further show that the method generalizes well to mismatched array geometries and exhibits strong robustness to corrupted microphone position metadata. Finally, we outline a natural extension of the approach to multi-source tracking and present the theoretical modifications required to support it.
\end{abstract}


\section{Introduction}

Sound source localization or tracking consists of estimating the position or direction of arrival of one or more sources based on signals recorded by a multi-microphone array. Traditionally, this is achieved using classical array signal processing methods, such as evaluating candidate directions \cite{schmidt1986music, dibiase2000srp}, analyzing system eigenstructures \cite{roy1989esprit}, or fitting parametric array models \cite{so2011maximumlikelihood}. These methods, however, exhibit several limitations, including the need for grid searches, precise array calibration, and a strong dependence on initialization.

With the growing interest in machine learning, neural network-based methods have been proposed as alternatives for the source localization and tracking problems using, e.g., multilayer perceptron architectures \cite{he2018mlps}, convolutional neural networks \cite{chakrabarty2017cnns}, and convolutional-recurrent networks \cite{adavanne2021doanetcrns}. The different structures also consider different input features and training objectives (classification or regression). Toward the state-of-the-art, we can find architectures that are more tailored to the problem of sound source tracking, either employing multi-dimensional convolutional layers \cite{diazguerra2021cross3d} or parallelizing its encoder for more efficient and general operation \cite{grinstein2024neuralsrp}. The neural network-based approaches use supervised training, i.e., they depend on precise ground-truth source position values. Obtaining such labels can be costly in both resources and time, also impeding training or fine-tuning to be done on-device.

Alternatively, unsupervised learning–based approaches could be used to bridge this gap and eliminate the need for labeled data, while potentially offering more precise estimates than classical methods. However, the existing approaches developed in this direction are not fully unsupervised, as they depend on clean speech data, forming a weakly supervised setup \cite{he2021multispeakerdoa}, or rely on a small set of labeled samples \cite{nozaki2025sourceaware}. Some approaches that are unsupervised require image or video modalities to facilitate tracking \cite{tanaka2021imagesound, lin2023contrastive}, which translates to additional sensors, like cameras \cite{lin2019novelhearing}, increasing the cost of audio/hearing devices. A different method that only depends on audio information \cite{huang2020timedomainunsupervised} nevertheless needs simulated acoustic transfer functions during training, which require full knowledge of the array geometry and source positions. Moreover, the approach incurs very high computational cost and a large number of parameters, due to per-direction/per-microphone autoencoder models with four 1024-unit fully connected layers, making it infeasible for most audio devices, and it is further limited by ad-hoc design choices.

In contrast, we propose a physics-guided variational model for single-source sound tracking. Our method is built upon a variational autoencoder structure, inheriting its underlying theoretical foundations. The physics-based decoder injects spatial information into the latent space via backpropagation, enabling unsupervised estimation of source directions. We compare the variational model against classical and state-of-the-art approaches across three experiments using real-world data, achieving performance comparable to supervised learning-based models. Furthermore, we demonstrate that our approach generalizes well to microphone array geometries that differ between training and testing. Compared to state-of-the-art methods, it also exhibits substantial robustness to corrupted metadata. Finally, we derive a multi-source extension of the proposed method and outline the theoretical framework required for its implementation. In this paper, however, we focus on the single-source case, as the multi-source implementation requires significant additional work and would shift the scope of the manuscript.

This paper is organized as follows. Section~\ref{sec:soundtracking} presents a literature review of traditional, supervised and unsupervised learning-based approaches for sound source tracking. In Section~\ref{sec:proposed} we present the proposed model, while detailing its architecture in Section~\ref{sec:architecture}. The experiments we perform, their results and discussion are shown in Section~\ref{sec:experiments}, with final conclusions and future work in Section~\ref{sec:conclusions}.

\section{Sound source tracking}
\label{sec:soundtracking}

Formally, to define the sound source tracking problem, we assume that $M$ microphones are positioned at $\mathbf{v}_m \in \mathbb{R}^3$, $m=1,...,M$, representing 3D Cartesian coordinates, and each microphone receives a signal frame of length $L$
\begin{equation}
    \mathbf{x}_m(t) = \sum_{n=1}^{N} \mathbf{h}_{nm}(t) \ast \mathbf{s}_n + \mathbf{\epsilon}_m(t),
\end{equation}
where ($\ast$) denotes the convolution between a room impulse response $\mathbf{h}_{nm}(t)$, from source $n$ to microphone $m$, and the signal produced by a source $n$ ($\mathbf{s}_n$). The number of sources $n$ is limited to $N$, and the sensor noise is denoted by $\mathbf{\epsilon}_m(t)$. The objective is to track the position of source $\mathbf{s}_n$, at discrete time $t$. Moreover, we assume that the sources are in the far-field of the array, i.e., the distance between microphones is significantly shorter than the distance between sources to the array \cite{grinstein2024srpreview}. The sound wave is thus treated as a plane wave without defined origin, and we only estimate its direction of arrival (DOA), described by an azimuth and an elevation angle. While this study restricts itself to far-field conditions, the underlying methodology is compatible with near-field extensions, which we leave for future work. Furthermore, we tackle the problem of tracking a single source ($N=1$), using unsupervised machine learning. Relevant existing methods related to the problem are explored next. 

\subsection{Classical}

Traditionally, sound source localization or tracking is performed via signal processing, in which various methods were proposed. The multiple signal classification (MUSIC) \cite{schmidt1986music} estimates the noise subspace from the sample covariance and sweep candidate arrival angles. MUSIC has a very high resolution but requires good calibration and model order choice. When sources are coherent, spatial smoothing is often necessary. Furthermore, the estimation of signal parameters via rotational invariance techniques (ESPRIT) \cite{roy1989esprit} uses the rotational invariance between identical, shifted sub arrays to get DOAs from a generalized eigenvalue problem, without grid search. ESPRIT is efficient and has a high resolution, but relies on precise array structure and calibration. Alternatively, maximum likelihood-based approaches \cite{so2011maximumlikelihood}, fit a parametric array model by maximizing the likelihood of the observations. Although very flexible, these methods are non-convex, computationally heavy, and very sensitive to initialization. Furthermore, careful regularization and constraints must be taken into account to avoid local minima.

In this work, we use the steered response power (SRP) method \cite{dibiase2000srp, grinstein2024srpreview} as a baseline, as it is a representative spatial signal-processing approach. SRP estimates the DOA by evaluating a predefined grid of candidate directions and assigning each direction a score that reflects how well the microphone signals align under that hypothesized geometry. It can be interpreted as a spatial projection of pairwise microphone cross-correlations, using the time delays predicted by each candidate location. Common features used within SRP include raw cross-correlation, generalized cross-correlation (GCC), and its weighted variant, the GCC with phase transform (GCC-PHAT). The latter normalizes the cross-spectrum by the magnitude, emphasizing phase information, which is more robust at low signal-to-noise ratios (SNRs). On the negative side, the resolution of SRP depends on the granularity of the direction grid and, in contrast to probabilistic approaches such as maximum likelihood, it does not model uncertainty.

\subsection{Supervised learning}

Various neural network-based methods using supervised learning were proposed for sound source localization and tracking using classification or regression \cite{perotin2019regressionvsclassification}. Multi-layer perceptrons (MLP) \cite{he2018mlps}, convolutional neural networks (CNNs) \cite{chakrabarty2017cnns}, convolutional-recurrent networks \cite{adavanne2021doanetcrns}, and graph neural networks \cite{grinstein2023graphs} are among the most used architectures for the task.

Among supervised approaches, two methods can be distinguished as representing the state of the art: Cross3D \cite{diazguerra2021cross3d} and Neural-SRP \cite{grinstein2024neuralsrp}. The Cross3D method was proposed for single-source DOA tracking, taking an SRP-produced map as input. Its architecture is a 3-dimensional CNN, where each dimension corresponds to elevation angle, azimuth angle, and time. The Neural-SRP model, on the other hand, is able to generalize with respect to both the number and the positions of the microphones in the array. It processes pairwise GCC-PHAT inputs using parallel copies of the same convolutional-recurrent encoder. The encoder outputs are summed and passed through an MLP, which produces the estimated DOA. Both methods are trained with simulated data generated using the image source method \cite{allen1979imagemethod} and tested on a real-world dataset.

\subsection{Unsupervised learning}
\label{ssec:unsupervised}

Acquiring data and defining precise direction labels can be an expensive, time-consuming process, leading some approaches to rely on weakly-labeled data \cite{he2021multispeakerdoa}, or to still rely on a small set of labels \cite{nozaki2025sourceaware}. Some methods also combine audio to image or video to facilitate sound source tracking \cite{tanaka2021imagesound, lin2023contrastive}. In contrast, some approaches have developed unsupervised adaptation methods for neural networks that are trained with synthetic data with labels. For example, \cite{takeda2017adaptation} adapts a pretrained model to unknown speaker positions and acoustic conditions. More recent work \cite{zhong2025importanceweighteddomainadaptationsound} proposes an unsupervised domain adaptation method to reduce the gap between synthetic training data (labeled) and real test data. 

We are, however, interested in fully unsupervised machine-learning-based methods that rely only on audio information for sound source localization or tracking. Such methods are, unfortunately, very scarce in the literature and exhibit strong limitations. To the best of our knowledge, the approach described next is the only prior work that is fully unsupervised. A time-domain unsupervised learning-based method is proposed in \cite{huang2020timedomainunsupervised}, using an autoencoder neural network. Multiple encoders recover a latent source signal, followed by paired decoders that model the forward acoustic transfer. The training loss is based on the reconstruction quality of the acoustic transfers and the cross-channel consistency among encoder outputs. This method, however, results in a very high computational cost and parameter count due to its multiple encoders and decoders, driven by the need to reconstruct sound source signals. Next, we describe our proposed solution, which adopts a more feasible, lower-complexity design.

\section{Proposed model}
\label{sec:proposed}

\begin{figure}[!t]
    \centering
    \includegraphics[width=0.55\linewidth]{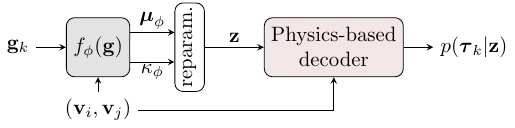}
    \caption{The proposed physics-guided variational model. Notice that $(\mathbf{v}_i,\mathbf{v}_j)$ is treated as metadata instead of a random variable.}
    \label{fig:vaesslmodel}
    \vspace{-4mm}
\end{figure}

The proposed model is composed of a variational encoder and a physics-based decoder, and is shown in Fig.~\ref{fig:vaesslmodel}. GCC-PHAT features $\mathbf{g}$ and the relative geometry of the microphone array $\mathbf{v}$ are fed to a variational encoder $f$ with parameters $\phi$, which outputs the mean $\boldsymbol{\mu}_{\phi}$ and concentration $\kappa_\phi$ of a von Mises-Fisher distribution. While $\mathbf{g}$ is treated as a random variable and part of the stochastic process (for being the input to the variational encoder), the relative microphone pair $k=(i,j)$ position to the array centroid $(\mathbf{v}_i,\mathbf{v}_j)$ is considered known and treated as metadata. The latent variable $\mathbf{z}$ is obtained through reparameterization. A physics-based decoder, which is only necessary during training, estimates pairwise sample time delays $\hat{\tau}_k$ and outputs a Gaussian likelihood $p(\boldsymbol{\tau}_k|\mathbf{z})$ centered at $\hat{\tau}_k$ -- which is then compared to an input distribution for loss calculation purposes. This sections describes the model parts and loss function.

\subsection{Input features}
\label{ssec:features}

We consider the generalized cross-correlation with phase transform as input features for microphone pair,
\begin{equation}
    \mathbf{g}_{k} = \mathrm{IDFT}\left( \frac{\mathbf{X}_i}{|\mathbf{X}_i|} \odot \frac{\mathbf{X}_j^*}{|\mathbf{X}_j|}  \right),
\label{eq:inputs}
\end{equation}
where $X_i$ and $X_j$ are, respectively, the discrete Fourier transform (DFT) of $\mathbf{x}_i$ and $\mathbf{x}_j$, and $\odot$ is the Hadamard product, with an $L$-sized inverse discrete Fourier transform (IDFT). The generated GCC-PHAT, $\mathbf{g} = [\mathbf{g}_{k = (i,j)}]$, for all $M(M-1)/2$ microphone pairs, has the shape of ($M(M-1)/2$, $T$, $G$), with $T$ time frames and $G$ central time-delay bins $\boldsymbol{\tau}_k$. We discard delay values where the absolute value exceeds the maximum theoretical time difference of arrival (TDOA), which can be obtained as
\begin{equation}
    |\tau_{max}| = \max \frac{||\mathbf{v}_i - \mathbf{v}_j||}{c} \ \forall \ (i,j) \in [1,M], \ i\neq j,
\end{equation}
where $c$ is the speed of sound. The GCC-PHAT features are input to a variational encoder, which is described next.

\subsection{Variational autoencoder}

We start by formulating the model as a variational autoencoder, with joint distribution given by
\begin{equation}
    p_\theta(\mathbf{g},\mathbf{z}) = p_\theta(\mathbf{g}|\mathbf{z})p(\mathbf{z}),
\end{equation}
with model parameters $\theta$ and a continuous latent variable $\mathbf{z}$. The model parameters could be obtained by evaluating the log-likelihood $\log p(\mathbf{g})$, however, requiring an intractable integral. 

To solve the intractability, we introduce a variational distribution with parameters $\phi$, $q_\phi(\mathbf{z}|\mathbf{g})$, that approximates the real posterior $p_\theta(\mathbf{z}|\mathbf{g})$. For any distribution $q_\phi(\mathbf{z}|\mathbf{g})$ such that $q_\phi(\mathbf{z}|\mathbf{g})>0$ wherever $p_\theta(\mathbf{g},\mathbf{z})$ for that given $\mathbf{g}$, we can write the marginal (log) likelihood as
\begin{equation}
    \log p_\theta(\mathbf{g}) = \log \int q_{\phi}(\mathbf{z}|\mathbf{g}) \frac{p_\theta(\mathbf{g},\mathbf{z})}{q_\phi(\mathbf{z}|\mathbf{g})} d \mathbf{z},
\label{eq:marg}
\end{equation}
where $\int q_{\phi}(\mathbf{z}|\mathbf{g}) [\cdot] d \mathbf{z}$ represents expectation of $[\cdot]$ over $q_{\phi}(\mathbf{z}|\mathbf{g})$, which can be used in  \eqref{eq:marg} as
\begin{equation}
    \log p_\theta(\mathbf{g}) = \log \mathbb{E}_{q_{\phi}(\mathbf{z}|\mathbf{g})} \left[  \frac{p_\theta(\mathbf{g},\mathbf{z})}{q_\phi(\mathbf{z}|\mathbf{g})} \right].
\label{eq:margwithexpect}
\end{equation}
Applying Jensen's inequality to \eqref{eq:margwithexpect} yields
\begin{equation}
    \log p_\theta(\mathbf{g}) \geq \mathbb{E}_{q_{\phi}(\mathbf{z}|\mathbf{g})} \left[ \log \frac{p_\theta(\mathbf{g},\mathbf{z})}{q_\phi(\mathbf{z}|\mathbf{g})} \right],
\end{equation}
which is defined as the evidence lower bound (ELBO) and right hand-side can be expanded as \cite{kingma2014autoencoding}
\begin{equation}
    \mathcal{L}_{\text{ELBO}}(\theta, \phi) =  \mathbb{E}_{q_{\phi}(\mathbf{z}|\mathbf{g})}[\log p_\theta(\mathbf{g}|\mathbf{z})]  - D_{\text{KL}}(q_\phi(\mathbf{z}|\mathbf{g})||p(\mathbf{z})).
\label{eq:elbo}
\end{equation}
The term $\mathbb{E}_{q_{\phi}(\mathbf{z}|\mathbf{g})}[\log p_\theta(\mathbf{g}|\mathbf{z})]$ is the reconstruction error and $D_{\text{KL}}(\cdot)$ is the Kullback-Leibler (KL) divergence between the variational distribution and the prior of $\mathbf{z}$. Notice that a hyperparameter $\beta$ can be applied to the KL divergence for better control over training \cite{higgins2017betavae}.

In our case, nonetheless, we substitute the decoder with a physics model, which imposes physical meaning to the latent $\mathbf{z}$ and, consequently, to the variational distribution. More specifically, we want to modify $p_\theta(\mathbf{g}|\mathbf{z})$ such that it depends only on physics aspects, avoiding trainable parameters that could result in a mismatch with the physical behavior.

\subsection{Physics-based decoder}

We can define the decoder output as the probability distribution $p(\boldsymbol{\tau}_k|\mathbf{z})$ over a discrete set of pairwise delay bins $\boldsymbol{\tau}_k$ spanning $[-\tau_{max}, \tau_{max}]$, which is computed from Gaussian-like logits centered at an estimated pairwise time-delay, $\hat{\tau}_k$.

First, we use the (latent variable) sampled position $\mathbf{z}$ to estimate the pairwise delays $\hat{\tau}_k$ from $\mathbf{z}$ to the microphone pairs $k$, as
\begin{equation}
    \hat{\tau}_{k}(\mathbf{z}) = \frac{(\mathbf{v}_i - \mathbf{v}_j)^T \mathbf{z}}{c} F_s,
\label{eq:tau}
\end{equation}
where $F_s$ is the sampling frequency. Then, we can calculate logits
\begin{equation}
    \boldsymbol{\ell}_k (\hat{\tau}_k) = -\frac{1}{2} \left( \frac{\boldsymbol{\tau}_k - \hat{\tau}_k}{\sigma} \right)^2,
\label{eq:logitsoutput}    
\end{equation}
which represent the unnormalized Gaussian log-probabilities over time-delay bins centered at the predicted $\hat{\tau}_k$. The standard deviation $\sigma$ is global since it reflects the overall uncertainty of the physics-based likelihood, rather than variability specific to individual microphone pairs. Finally, we apply a $\mathrm{softmax}$ over time-delay bins, yielding the normalized distribution
\begin{equation}
    p(\boldsymbol{\tau}_k|\mathbf{z}) = \frac{\exp(\boldsymbol{\ell}_k)}{\sum_{\boldsymbol{\tau}_k} \exp (\boldsymbol{\ell}_{k})},
\label{eq:est_dist}
\end{equation}
which approximates a (discretized) Gaussian centered at $\hat{\tau}_k$ with variance $\sigma^2$. The standard deviation $\sigma$ in \eqref{eq:logitsoutput} is a hyperparameter. t is important to note that the physics-based decoder is used exclusively during training. During inference, only the encoder is used.

\subsection{Variational posterior}

We consider the von Mises-Fisher (vMF) distribution for defining the variational posterior, which is a natural choice for directional statistics \cite{mardia1999Directional, ueno2021directionally}. Operating on the $n$-sphere, it can be used for modeling azimuth/elevation measurements \cite{garciafernandez2019vonmisesfisherdoa}. Specifically, a directional quantity $\mathbf{z}$ can be represented as a unit vector $\mathbf{z} \in \mathbb{S}^{n-1} = \{ \xi:\xi^T\xi=1, \ \xi\in \mathbb{R}^{n} \}$, i.e., $||\mathbf{z}|| = 1$. For a 3-dimensional problem, we set $n=3$. The vMF probability density in 3D is given by
\begin{equation}
    p(\mathbf{z}|\boldsymbol{\mu}_\phi,\kappa_\phi) = 
    \frac{\kappa_\phi}{4 \pi \sinh{\kappa_\phi}} \exp(\kappa_\phi \boldsymbol{\mu}_\phi^T \mathbf{z}),
\label{eq:variationalposterior}
\end{equation}
with $||\boldsymbol{\mu}_\phi|| = 1$, where the first term is a normalization constant. The use of the vMF posterior also constraints the azimuth and elevation angles to feasible values, as well as the norm of $\mathbf{z}$ to 1. The prior distribution of $\mathbf{z}$, on the other hand, is a uniform distribution derived from \eqref{eq:variationalposterior} with $\kappa_\phi = 0$, resulting in $p(\mathbf{z}) = \frac{1}{4\pi}$. The prior choice allows the encoder to consider the entire sphere, uniformly. Prior knowledge could be used to define a more specific prior.

Additionally, we consider $\boldsymbol{\mu}_\phi$ to be normal joint \textit{Cartesian} as a unit vector. If used in the angle form, discontinuities could occur at $\pm\pi$ (azimuth) and $\pm\pi/2$ (elevation), in the sense that both signs of $\pm\pi$ or $\pm\pi/2$ represent the same direction, but are on opposite ends of the interval. For directional of arrival estimation in inference mode, $\boldsymbol{\mu}_\phi$ is converted to angular dimensions (elevation/azimuth) by considering a unitary radius and is taken as the estimated DOA. During training, on the other hand, the reparameterization of the vMF distribution is needed and described below.

\subsection{Reparameterization}

We adapt the reparameterization trick from \cite{kingma2014autoencoding} to the vMF in \eqref{eq:variationalposterior} by sampling through a canonical frame and then rotating to the target mean \cite{Ulrich1984vonmisesreparam}. Concretely, we draw two independent uniform variables $u_1,u_2 \sim \mathcal{U}(0,1)$. The first controls the elevation component $w$ via the inverse cumulative distribution function (CDF) of the canonical vMF (mean at the north pole), whose CDF for $w\in[-1,1]$ is
\begin{equation}
    F(w)=\frac{\big(e^{\kappa_\phi w}-e^{- \kappa_\phi}\big)}{\big(e^{\kappa_\phi}-e^{-\kappa_\phi}\big)}.
\label{eq:cdf}
\end{equation}
By inverting \eqref{eq:cdf}, we obtain

\begin{equation}
    w = \frac{1}{\kappa_\phi}\,\log\!\big((1-u_1)e^{-\kappa_\phi}+u_1 e^{\kappa_\phi}\big),
\label{eq:inv_cdf}
\end{equation}
with the isotropic limit $w=2u_1-1$ as $\kappa_\phi\to 0$ -- uniform on $[-1,1]$. 

The second uniform controls the azimuth by making it equivalent to $2\pi u_2$, and we set the tangential radius $r=\sqrt{1-w^2}$. We can construct a local orthonormal basis $(\mathbf{e}_1,\mathbf{e}_2,\boldsymbol{\mu}_\phi)$ by picking a reference axis $\mathbf{r}$ not collinear with $\boldsymbol{\mu}_\phi$, and define
\begin{equation}
    \mathbf{e}_1 = \frac{\mathbf{r}\times \boldsymbol{\mu}_\phi}{\|\mathbf{r}\times \boldsymbol{\mu}_\phi\|}, 
    \qquad
    \mathbf{e}_2 = \boldsymbol{\mu}_\phi \times \mathbf{u}.
\label{eq:orthonormal_basis}
\end{equation}
A canonical sample on $\mathbb{S}^2$ is then mapped to the $\boldsymbol{\mu}_\phi$ frame as
\begin{equation}
    \mathbf{z} = r\cos(2\pi u_2) \mathbf{u} + r\sin(2\pi u_2) \mathbf{v} + w \boldsymbol{\mu}_\phi,
\end{equation}
with $||z|| = 1$. 

\subsection{Loss function}

For a proper unsupervised operation, we consider \eqref{eq:elbo} as the basis for the loss function. However, given the use of a physics-based decoder that does not literally reconstructs the input signal, we modify the first part of the ELBO. As the decoder estimates a probability distribution over pairwise delay bins, $p(\boldsymbol{\tau}_k|\mathbf{z})$, we first normalize the inputs \eqref{eq:inputs} as
\begin{equation}
    \tilde{\mathbf{g}}_k = \frac{\mathbf{g}_k - \boldsymbol{\mu}(\mathbf{g}_k)}{\boldsymbol{\Sigma}(\mathbf{g}_k) + \epsilon},
\label{eq:inputlogits}
\end{equation}
with mean
\begin{equation}
    \boldsymbol{\mu}(\mathbf{g}_k) = \frac{1}{G} \sum_{\boldsymbol{\tau}_k} \mathbf{g}_k    
\end{equation}
and variance
\begin{equation}
    \boldsymbol{\Sigma}(\mathbf{g}_k) = \frac{1}{G} \sum_{\boldsymbol{\tau}_k} (\mathbf{g}_k - \boldsymbol{\mu}(\mathbf{g}_k))^2.
\end{equation}
Hence, we can obtain the input time-delay distribution by applying a weighted $\mathrm{softmax}$ function:
\begin{equation}
    p(\boldsymbol{\tau}_k|\mathbf{g}_k) = \frac{\exp (\lambda  \tilde{\mathbf{g}}_k)}{\sum_{\boldsymbol{\tau}_k} \exp ( \lambda  \tilde{\mathbf{g}}_k)},
\label{eq:input_dist}
\end{equation}
with $\lambda$ being a hyperparameter to scale the GCC distribution. 

The physics loss can be calculated as the KL divergence between the estimated time-delay distribution \eqref{eq:est_dist} and the input time-delay distribution \eqref{eq:input_dist} as
\begin{equation}
    \mathcal{L}_{\text{physics}} = \mathbb{E} \left[\mathbf{a}_\mathbf{x} \sum_{k} D_{KL}\left( p(\boldsymbol{\tau}_k|\mathbf{g}_k) \ || \ p(\boldsymbol{\tau}_k|\mathbf{z}) \right) \right], 
\end{equation}
which can be approximated by the cross entropy
\begin{equation}
    \mathcal{L}_{\text{physics}} = \mathbb{E} \left[
        - \mathbf{a}_\mathbf{x} \sum_{k} \sum_{\boldsymbol{\tau}_k}
        p(\boldsymbol{\tau}_k|\mathbf{g}_k)
        \log p(\boldsymbol{\tau}_k|\mathbf{z})
    \right].
\label{eq:physicsloss}
\end{equation}
We also apply a sound activity mask $\mathbf{a}_{\mathbf{x}}$ on the physics loss, relating to the time-axis of the microphones' outputs $\mathbf{x}$, avoiding unpredictable behavior when the system would estimate a position during silence. This was previously observed as necessary for VAE applications with audio signals \cite{fiorio2025clustering}. The activity mask can be obtained via simulation or estimated with existing methods \cite{jongseo1999vad}, which does not violate the unsupervised learning nature of the proposed approach as it can be estimated purely based on microphone sensor signals. For demonstration purposes in this paper, however, we use the true activity mask provided by the datasets in the training of all models. Note that \eqref{eq:physicsloss} computes the expected negative log-likelihood of the estimated distribution under the input distribution.

Moreover, we also regularize the encoder by the KL divergence between the vMF distribution given estimated encoder parameters and a uniform distribution in the sphere $\mathbb{S}^2$. In \eqref{eq:elbo}, the KL regularization loss, defined in the ELBO as 
\begin{equation}
    \mathcal{L}_{\mathrm{KL}} = D_{\mathrm{KL}} \left( q_\phi(\mathbf{z}|\mathbf{g})\ ||\ p(\mathbf{z}) \right),
\end{equation}
can be calculated in closed form with the aforementioned considerations as
\begin{equation}
    \mathcal{L}_{\mathrm{KL}} = D_{\mathrm{KL}} \left( \mathrm{vMF}(\boldsymbol{\mu}_\phi, \kappa_\phi)\ ||\ \mathrm{Uniform}(\mathbb{S}^2) \right) = \kappa_\phi\left( \coth{\kappa_\phi} - \frac{1}{\kappa_\phi} \right) + \log \kappa_\phi - \log \sinh{\kappa_\phi}.
\label{eq:klregloss}
\end{equation}

Finally, we can write the modified ELBO for the directional of arrival estimation with the physics-guided variational encoder,
\begin{equation}
    \mathcal{L}_{\text{ELBO},\text{DOA}}(\theta, \phi) =  \mathcal{L}_{\text{physics}} - \beta \mathcal{L}_{\mathrm{KL}},
\label{eq:doaelbo}
\end{equation}
by combining both physics \eqref{eq:physicsloss} and KL regularization \eqref{eq:klregloss} loss terms, with the hyperparameter $\beta$ for more stable training \cite{higgins2017betavae}.

\section{Architecture}
\label{sec:architecture}

\begin{figure*}[!t]
\centering
    \begin{subfigure}{.99\textwidth}
        \centering
        \includegraphics[width=0.55\textwidth]{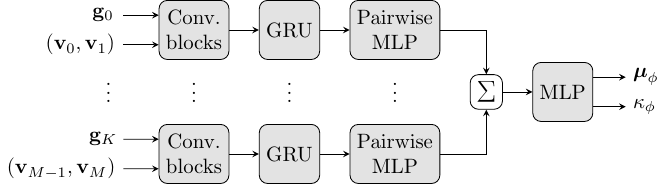}
        \caption{Encoder}
        \label{fig:VAE-M2}
    \end{subfigure} \\ \vspace{3mm}
    \begin{subfigure}{.99\textwidth}
        \centering
        \includegraphics[width=0.625\textwidth]{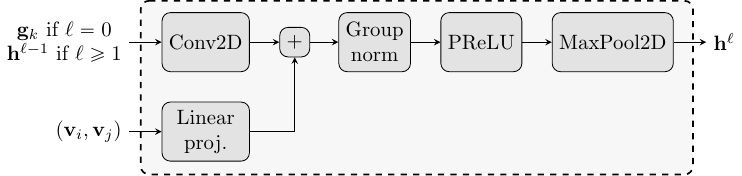}
        \caption{Convolutional block, where $\ell$ indicates the convolutional block number}
        \label{fig:VAE}
    \end{subfigure} 
\caption{Encoder architecture. A conv. block takes input $\mathbf{x}$ and metadata $\mathbf{v}$, and is composed of a 128-channel output 2D conv. layer, with kernel (3,3), unitary stride and padding. Its output is biased by a metadata projection through a linear layer. Single-group normalization is applied, which output passes through a PReLU and a 2D max pooling layer. The encoder is composed of three conv. blocks, with pooling sizes (2,2,2) in the lag axis and (5,1,1) in the time axis. They are followed by a 128-sized output unidirectional gate recurrent unit (GRU) with two layers, and a pairwise MLP of two layers, both activated by PReLU with output of size 128. The pairwise MLP outputs are combined by summation, which is then processed by a final MLP of two layers. The first layer has a PReLU activation and the second is purely linear, resulting in four outputs, of which three are the coordinates of $\boldsymbol{\mu}_\phi$, and the remaining represents $\kappa_\phi$.}
\label{fig:architecture}
\vspace{-4mm}
\end{figure*}

We consider a pairwise architecture where one encoder is considered per microphone pair. For a a reduction in number of parameters and efficient parallel processing, we share encoder weights, also allowing for a convenient scaling of number of microphones. The encoder architecture is shown in Figure~\ref{fig:architecture}. While the Neural-SRP \cite{grinstein2024neuralsrp} applied metadata (relative position of the microphone pair) after recurrent layers in the encoder, we apply the metadata $(\mathbf{v}_j,\mathbf{v}_j)$ to the encoder in each of the convolutional blocks, after the convolutional layers. This represents an early bias to the encoder that accelerates convergence and makes the system aware of the microphone relative positions, which is of substantial importance to the direction-related variables it is estimating. More information on the use of metadata can be seen in \cite{Grinstein2023DualInput}.

Similarly to the Neural-SRP \cite[Section~IV]{grinstein2024neuralsrp}, we use a (3,3) kernel in the 2D convolutions, which operates both in the lag and the time axis. While convolutions can be avoided in the time axis for many supervised tasks, notes on the application of VAEs for audio \cite{fiorio2025clustering} have shown that a non-unitary kernel in the time axis can dramatically enhance the performance. Importantly, with a non-unitary kernel in the time axis and pooling, the latent space time axis is of different size than that of the input. To properly work with two different time axis, we linearly interpolate both the sound activity mask $\mathbf{a}_\mathbf{x}$ time axis and the GCC-PHAT features $\mathbf{g}$ before obtaining the probability distribution over time delays for input features \eqref{eq:inputlogits}.

\section{Experiments}
\label{sec:experiments}

We evaluate the proposed model through three experiments which are based in three different training settings. For all cases, the training data is obtained synthetically. Crucially, the data used in testing comes from a real-world setting, allowing us to see how well the unsupervised method can generalize to unseen data.

\subsection{Data}
\label{ssec:data}

All training data is based on recordings, of a compact stationary microphone array, with a single moving source inside a cuboid-shaped reverberant room, following a 3-dimensional sinusoidal trajectory. The trajectories as generated by randomly selecting start and end points in the room. A 3-dimensional sinusoidal vector with random amplitude and frequency is assigned for each trajectory. The acoustic scenarios are randomly generated using the image source method \cite{diazguerra2021cross3d,allen1979imagemethod}. We follow similar simulation parameters to that of \cite{grinstein2024neuralsrp}, presented in Table~\ref{tab:common parameters}. All simulations are executed with the gpuRIR python package \cite{DiazGuerra2021gpuRIR}.

\begin{table}[!t]
    \centering
    \begin{tabular}{c c c}
        \hline
        \textbf{Parameter} & \textbf{Min. value} & \textbf{Max. value} \\
        \hline 
        R60 (s) & 0.2 & 1.0 \\
        SNR (dB) & 5 & 30 \\
        Oscillations & 0 & 2 \\
        Oscillations amp. (m) & 0 & 1 \\
        \hline
    \end{tabular} 
    \vspace{1mm}
    \caption{Training data common variable parameters.}
    \label{tab:common parameters}
    \vspace{-8mm}
\end{table}

For clean speech data, we use the LibriSpeech dataset \cite{panayotov2015librispeech}, more specifically the train-100-clean set, and the test-clean set for validation. Each training or validation data sample is composed of a 20 seconds audio trace, with 16 kHz sampling rate, where multiple, randomly chosen, utterances from a same chapter in the LibriSpeech dataset are concatenated together. The total length of the dataset is equivalent to the number of chapters in LibriSpeech. 

For each considered experiment, we have a different setting on the training and validation data. This is described in the following. \emph{Experiment 1} contains additive white Gaussian noise (AWGN) in the training and validation data, directly in the auralized signals at the desired SNR. This represents a situation where the predominant noise comes from the microphones or related hardware. The considered microphone geometry matches the one in testing. In \emph{Experiment 2}, the array geometry also matches testing, and a second source emitting random Gaussian noise is randomly placed at a static position in the room. In this case, the noise is considered to come from the environment. Finally, \emph{Experiment 3} deals with uncertainty in the microphones' position, where the position of each microphone is sampled from a Gaussian distribution centered at the test geometry, with standard deviation of 1 cm (roughly 10\% of the longest distance between two microphones in the array). Auralized AWGN is considered for this experiment.

For all cases, the LOCATA dataset \cite{evers2020locata} with a pseudo-spherical array geometry, 12 microphones, and a single source is considered for testing. The dataset consists of six different tasks with different complexity. Similarly to \cite{grinstein2024neuralsrp}, we select three of them -- static source, moving source, and moving microphone position -- which we average the results over. All data from LOCATA is resampled to 16 kHz.

To obtain the GCC-PHAT input features as described in Section~\ref{sec:proposed}-\ref{ssec:features} for both training, validation, and testing, we take the short-term Fourier transform of the time-domain signals with a window size of 4096 samples and a hop rate of 0.75. We choose $G = 64$ lags for the GCC-PHAT features, which are bounded by the theoretical maximum time-delay of arrival with the considered microphone geometry. Please refer to \cite[Section 2.3]{grinstein2024srpreview} for a detailed calculation.

\subsection{Baselines}

We consider both classical and supervised learning-based methods for comparison with the proposed unsupervised model. Due to elevated complexity with multiple encoders, we do not consider the prior unsupervised learning-based approach from \cite{huang2020timedomainunsupervised}, as it does not represent a feasible approach for most hardware-constrained audio processing devices.

The first baseline is the SRP with GCC-PHAT features, i.e., SRP-PHAT \cite{dibiase2000srp, grinstein2024srpreview}. We implement the SRP with a rectangular grid of 64 $\times$ 32, with first and second dimensions representing, respectively, azimuth and elevation. The implementation is done according to the time-domain formulation of SRP in \cite[Section~II]{grinstein2024srpreview}. For supervised-learning baselines, the Cross3D is considered and implemented according to \cite[Section~IV-B]{diazguerra2021cross3d} and take as input the SRP map obtained with the aforementioned SRP-PHAT baseline. We also take the Neural-SRP into account, implemented as \cite[Section~III-B]{grinstein2024neuralsrp}. For evaluation of all models, we use the root mean square angular error from \cite[Section~IV-A]{grinstein2024neuralsrp}.

\subsection{Hyperparameters}
\label{ssec:hyperparameters}

Hyperparameters are very influential for the performance of unsupervised methods \cite{higgins2017betavae}. Thereby, we specifically analyze the limits of $\sigma$, $\lambda$, and $\beta$, which serves as base to choose its values. From \eqref{eq:logitsoutput}, a standard deviation $\sigma \rightarrow 0$ (no uncertainty modeling) would cause the Gaussian to become extremely sharp around the predicted delay $\hat{\tau}_k$, and the softmax would assign almost all probability mass to that bin. Considering $\mathbf{z}$ a candidate position from a predefined grid would be equivalent to computing alignment scores for each candidate and picking the best one. This behavior is analogous to the classical SRP, i.e., our method is lower-bounded by the SRP performance with GCC-PHAT inputs. On the other hand, a $\sigma \rightarrow \infty$ would create a uniform distribution without any physics modeling. We choose $\sigma$ trainable considering that a variational autoencoder needs a parameterized encoder and decoder for proper training \cite{kingma2014autoencoding}. Additionally, we use a softplus function to guarantee that $\sigma$ remains positive.

In the limit, an input distribution scaling $\lambda \rightarrow \infty$ makes $p(\boldsymbol{\tau}_k|\mathbf{g}_k)$ from \eqref{eq:input_dist} a delta at the GCC-PHAT peak, and the physics-based loss reduces to the negative log-likelihood at that peak delay. With a Gaussian likelihood assumption, this is equivalent to training the model by taking the mean squared error of the estimated peak delays against the (input) GCC-PHAT peak delays, e.g., using a decoder that only estimates $\hat{\tau}_k$ from \eqref{eq:tau}. In other words, by choosing $\lambda \rightarrow \infty$, we would ignore the noise and reverberation present in the acoustic scene, which may worsen the performance as reflections might change the correct value of DOA. However, in some cases when omnidirectional/auralized (uniformly distributed) noise is added at low SNR, the input distribution might get closer to uniform, not presenting a clear peak. For this reason, a $\lambda > 1.0$ is suggested, with exact value to be determined on a case-by-case basis. Based on preliminary tests with the considered datasets and operating conditions, we choose $\lambda = 8.0$ as it achieved satisfactory performance. An optimal choice for $\lambda$ is outside the scope of this work.

Furthermore, we observed that the hyperparameter $\beta$, weighting the loss function \eqref{eq:elbo} and \eqref{eq:doaelbo}, can help to avoid a collapse of the latent space, as discussed in \cite{higgins2017betavae}. We implement a warm-up schedule by setting $\beta=0$ during the first 5\% of training epochs, with $\mathbf{z}=\boldsymbol{\mu}_\phi$, preventing an early posterior collapse. The remaining epochs assume a value of $\beta = 1.0$. This choice also relates to the rate-distortion tradeoff \cite{Alemi2018FixingBrokenELBO}, initially minimizing the distortion (reconstruction) to later achieve a certain rate (KL). 

The proposed method is trained for 300 epochs with a learning rate varying exponentially from 5e-4 to 5e-5 over epochs. Both baseline methods are trained for the amount of epochs mentioned in the original papers (80 epochs) and their originally proposed fixed learning rate of 1e-4. Note that a more extensive training is often needed with generative approaches. The batch size for all cases is 16. 

\begin{table}[!t]
    \centering
    \begin{tabular}{c c c c}
        \hline
        \textbf{Method} & \textbf{Learning type} & \textbf{Simulated data} & \textbf{LOCATA} \\
        \hline 
        SRP-PHAT & N.A. & 8.4 & 16.7 \\
        \hline
        Proposed & Unsupervised & 5.5 & 8.8 \\
        \hline
        Cross3D & Supervised & 4.2 & 6.1 \\
        Neural-SRP & Supervised & 3.2 & 4.7 \\
        \hline
    \end{tabular}     \vspace{1mm}
    \caption{Root mean square angular errors for Experiment 1 -- models trained with auralized AWGN and matching array geometries to that of testing. Proposed method evaluated over 10 independent runs (average); Cross3D and NeuralSRP results from \cite{grinstein2024neuralsrp}.}
    \label{tab:experiment1_results}
    \vspace{-4mm}
\end{table}

\begin{table}[!t]
    \centering
    \begin{tabular}{c c c c}
        \hline
        \textbf{Method} & \textbf{Learning type} & \textbf{Simulated data} & \textbf{LOCATA} \\
        \hline 
        SRP-PHAT & N.A. & 7.9 & 16.7 \\
        \hline
        Proposed & Unsupervised & 3.8 & 8.0 \\
        \hline
        Cross3D & Supervised & 3.5 & 6.1 \\
        Neural-SRP & Supervised & 3.4 & 5.8 \\
        \hline
    \end{tabular}     \vspace{1mm}
    \caption{Root mean square angular errors for Experiment 2 -- models trained with directional AWGN and matching array geometries to that of testing. Proposed method evaluated over 10 independent runs (average); Cross3D and NeuralSRP results from \cite{grinstein2024neuralsrp}.}
    \label{tab:experiment2_results}
    \vspace{-6mm}
\end{table}

\subsection{Results}

The root mean square angular errors for Experiments 1, 2, and 3, are shown, respectively, in Tables \ref{tab:experiment1_results}, \ref{tab:experiment2_results}, and \ref{tab:experiment3_results}, for all considered methods. For Experiments 1 and 2, the Cross3D and Neural-SRP results are extracted from \cite{grinstein2024neuralsrp}, as the experiments are defined in the same way as in the mentioned paper. 

When the models are trained with auralized AWGN (Experiment 1), we observe that the proposed unsupervised approach outperforms the SRP-PHAT and is outperformed by the supervised-learning models of Cross3D and Neural-SRP, for both simulated and real-world data. This is an expected order of performance, since the classical approach is limited by not modeling uncertainty nor being trainable, and the methods with supervised learning had access to the ground-truth values of direction of arrival. 

In Experiment 2, when considering directional AWGN in the signals representing ambient noise, a similar behavior can be observed. In this case, however, the proposed variational model could further improve the DOA estimation with the LOCATA dataset. This is probably due to a better correlation between simulated and real-world noise, as in real-world scenarios the noise is mostly directional, produced by a source located in space (environment noise). 

Conversely, we observe a different behavior in Experiment 3. By sampling the microphone positions from a Gaussian distribution around the LOCATA NAO robot’s array, we are able to evaluate the effect of uncertain microphone positioning on DOA estimation, i.e., training with metadata or data that differ from those used in evaluation. In this experiment, the proposed unsupervised approach is able to outperform the Cross3D method when evaluated on real-world data, achieving performance comparable to Neural-SRP. While both the proposed model and Neural-SRP rely on GCC-PHAT inputs and have access to metadata, the Cross3D method depends on an SRP map. Its reduction in performance likely occurs because it inherits limitations from the SRP method -- such as grid dependence -- which create a biased input that is further degraded by uncertain microphone positions. Cross3D also appears to have a lower capacity to generalize, given its substantially higher error on the LOCATA dataset compared to simulated data.

\begin{table}[!t]
    \centering
    \begin{tabular}{c c c c}
        \hline
        \textbf{Method} & \textbf{Learning type} & \textbf{Simulated data} & \textbf{LOCATA} \\
        \hline 
        SRP-PHAT & N.A. & 8.4 & 16.7 \\
        \hline
        Proposed & Unsupervised & 6.3 & 8.2 \\
        \hline
        Cross3D & Supervised & 6.7 & 11.6 \\
        Neural-SRP & Supervised & 6.1 & 7.6 \\
        \hline
    \end{tabular}     \vspace{1mm}
    \caption{Root mean square angular errors for Experiment 3 -- models trained with auralized AWGN, with training geometry that differs from evaluation. The models are evaluated for the correct geometry over 10 independent runs (average).}
    \label{tab:experiment3_results}
    \vspace{-6mm}
\end{table}

\begin{table}[!t]
    \centering
    \begin{tabular}{c c c c}
        \hline
        $(\mathbf{v}_i,\mathbf{v}_j)$ \textbf{corruption} (\%) & \textbf{Method} & \textbf{LOCATA} & \textbf{Increase} (\%)\\
        \hline
        \multirow{2}{*}{10} & Proposed & 8.7 & 4.6 \\
        & Neural-SRP & \textbf{8.0} & 5.3 \\
        \hline
        \multirow{2}{*}{20} & Proposed & 9.9 & 19.8 \\
        & Neural-SRP & \textbf{9.6} & 26.4 \\        
        \hline
        \multirow{2}{*}{30} & Proposed & \textbf{11.8} & 42.4 \\
        & Neural-SRP & 12.1 & 59.6 \\   
        \hline
        \multirow{2}{*}{40} & Proposed & \textbf{14.1} & 69.5 \\
        & Neural-SRP & 15.3 & 101.2 \\   
        \hline
        \multirow{2}{*}{50} & Proposed & \textbf{16.5} & 99.0 \\
        & Neural-SRP & 18.8 & 147.8 \\
        \hline
    \end{tabular}     \vspace{1mm}
    \caption{Root mean square angular errors for models trained as in Experiment 3, evaluated with corrupted metadata $(\mathbf{v}_i,\mathbf{v}_j)$, and percentage increase in relation to the baseline (Table~\ref{tab:experiment3_results}). Evaluation for 10 independent runs (average). The lowest error is highlighted for each case.}
    \label{tab:experiment35_results}
    \vspace{-8mm}
\end{table}

Furthermore, the real-world performance of the proposed variational model improves relative to the other experiments. This counterintuitive behavior may arise from the additional excitation introduced by varying microphone positions -- similar to system identification problems \cite{ljung1999system} -- combined with the uncertainty modeling inherent to the probabilistic latent space (via reparameterization) and the decoder (through probabilistic modeling of time delays). For methods lacking explicit uncertainty modeling (Cross3D and Neural-SRP), performance worsens for both simulated and real-world data.

As a direct comparison between the proposed unsupervised approach and Neural-SRP, we corrupt the metadata during evaluation at different levels. By applying AWGN to the metadata $(\mathbf{v}_i, \mathbf{v}_j)$ fed to the model, we cause it to differ from the real microphone positions. In practice, this is equivalent to a scenario where the system undergoes an initial calibration to define metadata values, but the microphone positions in the array are later changed without recalibration. We define the corruption percentage as the percentage of the maximum absolute value of $\mathbf{v}$ by which the unit-variance AWGN is scaled. From Table~\ref{tab:experiment35_results}, we see that as metadata corruption increases, the root-mean-square angular error of the proposed approach grows much less than that of the Neural-SRP model, and the unsupervised approach outperforms Neural-SRP from the 30\% corruption level onward. A plausible explanation is that the proposed approach is based on variational autoencoders, inheriting their generalization properties through modeling the underlying data distribution within a known latent distribution.

\begin{figure}
    \centering
    \includegraphics[width=0.55\linewidth]{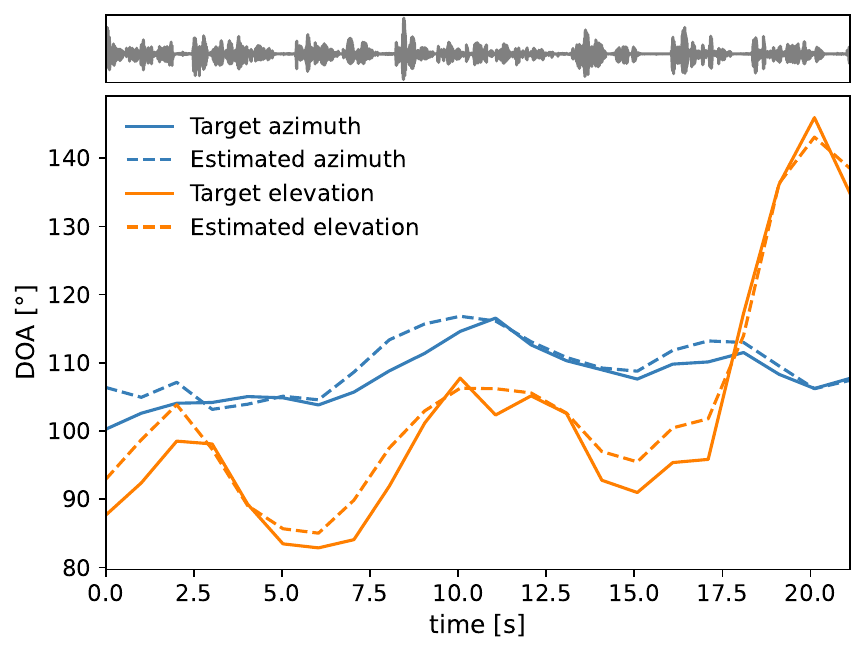}
    \caption{DOA estimation example of the proposed variational model for real-world (LOCATA) data. The time-domain signal captured by one of the microphones is displayed in gray, while blue and orange curves represent, respectively, the azimuth and elevation targets, with their dashed counterparts showing estimated angles.}
    \label{fig:doarandom}
\end{figure}

To facilitate visualization of the DOA estimation, in Figure~\ref{fig:doarandom} we display an example of elevation and azimuth angles estimated with the proposed variational model for 20-seconds of data from the LOCATA dataset. The model used for this example was trained as described in Experiment 1, achieving an average root mean square angle error of 3.7° for this excerpt.

\begin{table}[!t]
    \centering
    \begin{tabular}{c c c c}
        \hline
        \textbf{Method} & \textbf{Param.} ($\times 10^6$) & \textbf{MACs} ($\times 10^{9}$) & \textbf{RTF} \\
        \hline
        Proposed & 0.89 & 159 & 0.18 \\
        Cross3D & 5.63 & 41 & 0.01 \\
        Neural-SRP & 0.89 & 159 & 0.17 \\
        \hline
    \end{tabular}     \vspace{1mm}
    \caption{Number of parameters, MACs, and RTF obtained for the considered neural network-based methods.}
    \label{tab:complexity}
    \vspace{-4mm}
\end{table}

\subsection{Complexity analysis}
\label{sec:conclusions}
We evaluate the complexity of the proposed model in terms of parameters, multiply-accumulate operations (MACs), and real time factor (RTF), which is shown in Table~\ref{tab:complexity}. Importantly, the decoder is not taken into account for the proposed model as it is unnecessary during inference mode. Also note that we obtain the RTF by dividing the calculation time for one step (hop length) by the hop length itself, averaged over 2000 seconds of audio. Since the proposed model has a similar architecture to that of the Neural-SRP, its number of parameters, MACs, and RTF are very similar to the supervised baseline. The Cross3D, on the other hand, presents much more parameters but less MACs and RTF, which happens because the model does not include any recurrent layer, mainly convolutional layers.

\subsection{Posterior uncertainty analysis}

\begin{figure*}[!t]
\centering
    \begin{subfigure}{.32\textwidth}
        \centering
        \includegraphics[width=1\textwidth]{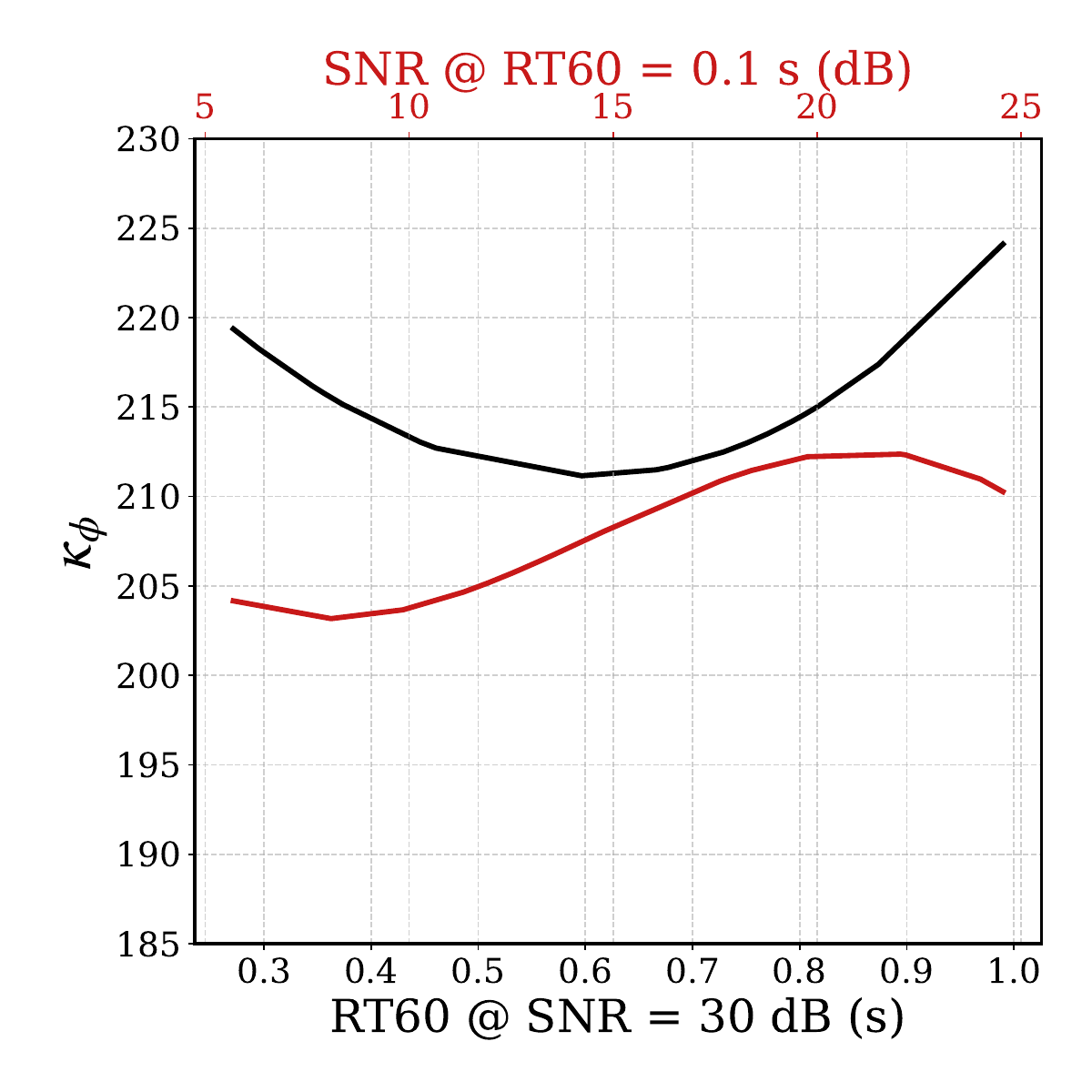}  
        \caption{Fixed RT60 of 0.1 s, SNR of 30 dB}
        \label{fig:tau_data}
    \end{subfigure}
    \begin{subfigure}{.32\textwidth}
        \centering
        \includegraphics[width=1\textwidth]{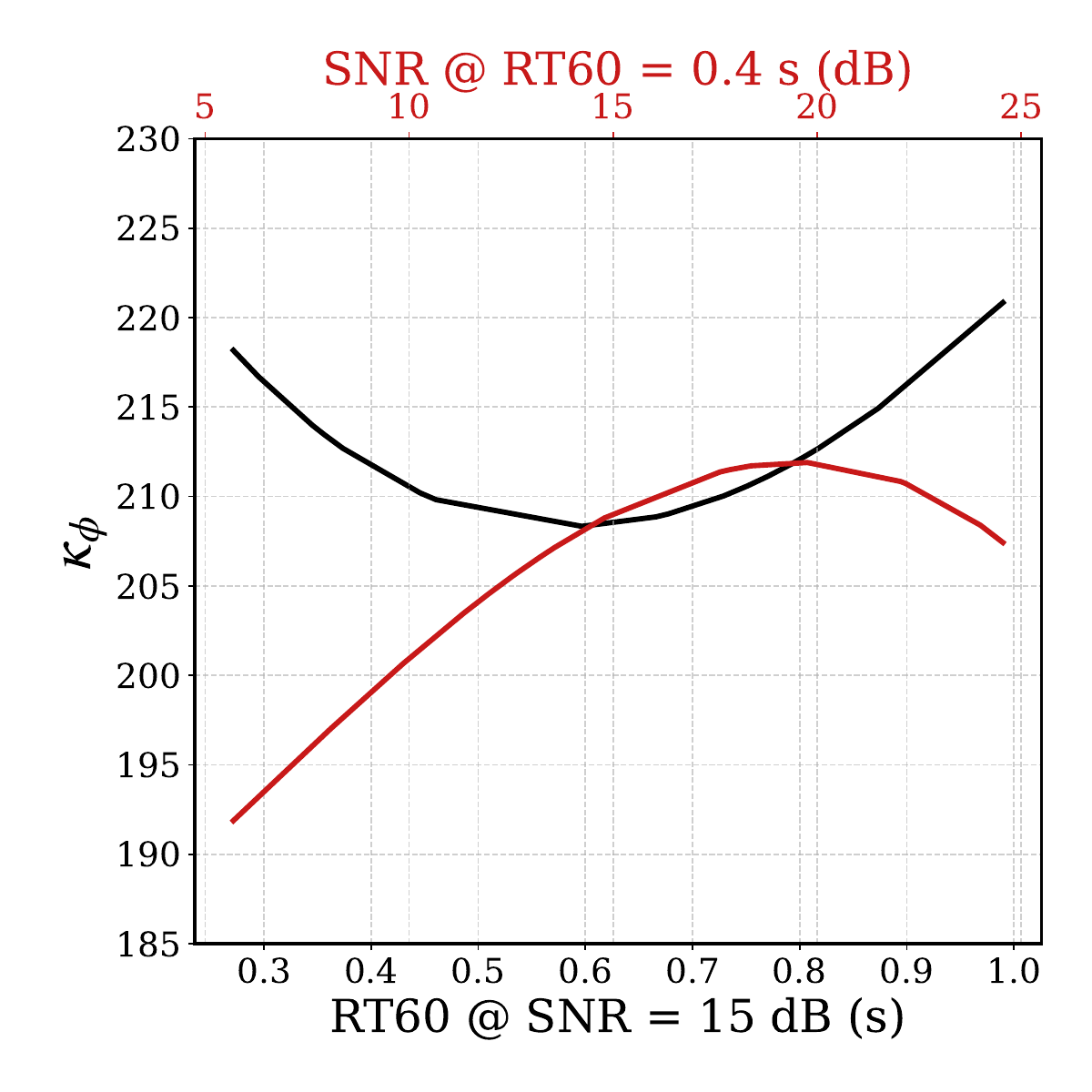}
        \caption{Fixed RT60 of 0.4 s, SNR of 15 dB}
        \label{fig:tau5_clusters}
    \end{subfigure} 
    \begin{subfigure}{.32\textwidth}
        \centering
        \includegraphics[width=1\textwidth]{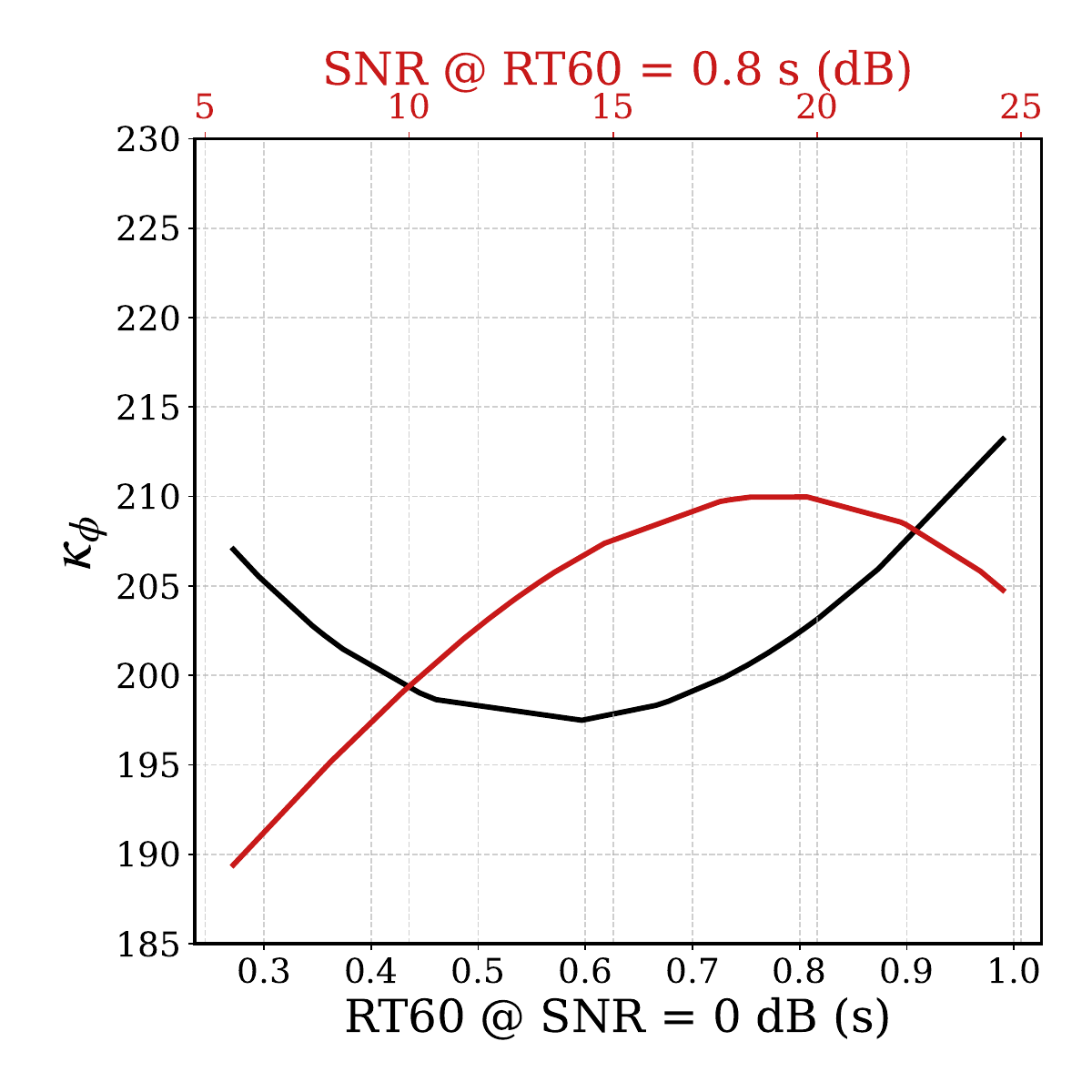}
        \caption{Fixed RT60 of 0.8 s, SNR of 0 dB}
        \label{fig:tau10_clusters}
    \end{subfigure}  \vspace{2mm}
\caption{Concentration $\kappa_\phi$ obtained with one model trained as in Experiment 1, with (upper axis, red) an SNR sweep at fixed RT60, and (lower axis, black) an RT60 sweep at fixed SNR, for different fixed conditions over the simulated test set.}
\label{fig:kappa}
\vspace{-4mm}
\end{figure*}

In Figure~\ref{fig:kappa}, we examine the behavior of the posterior concentration parameter under different reverberation and noise conditions. When the SNR is fixed and RT60 is varied (red curves), $\kappa_\phi$ shows a clear non-monotonic trend. For short RT60 values, the direct path dominates the GCC-PHAT structure, producing a higher peak in $p(\tau_k|\mathbf{g}_k)$ and resulting in larger $\kappa_\phi$, i.e., higher confidence in the inferred direction. As RT60 increases to moderate values, early reflections introduce competing peaks around the direct-path delay, making the time-delay distribution less clearly uni-modal. Since the physics-guided decoder models each pairwise delay with a uni-modal likelihood, the encoder reflects this increased ambiguity by reducing the concentration value. At high RT60, the reverberant field becomes more diffuse and the structured secondary peaks weaken, causing the dominant mode to stabilize again. In this regime, $\kappa_\phi$ increases slightly, indicating that the model recovers part of its directional certainty despite the overall broader input distributions.

On the other hand, when RT60 is fixed and SNR is varied (black curves), the behavior of the concentration parameter becomes more monotonic. At high SNR, the inter-microphone phase structure is preserved, yielding sharp GCC-PHAT peaks and larger concentration values. As SNR decreases, additive noise progressively masks the phase information and flattens the delay distribution, reducing the contrast between the direct path and surrounding delays. The posterior then widens to remain consistent with the physics-guided likelihood, producing steadily smaller $\kappa_\phi$. Unlike the reverberation case, low SNR does not create structured competing peaks, simply decreasing signal coherence. As a result, the decline in concentration with SNR is smooth and directly reflects noise-induced uncertainty in the latent DOA estimate.

\section{Multi-source extension guidelines}
\label{sec:multi}
Although in this paper we focus on the single-source case, the proposed method can be naturally extended to the multi-source setting. In an environment with $N$ sources, we model the latent direction as a mixture of von Mises–Fisher (vMF) distributions on $\mathbb{S}^2$:
\begin{equation}
q_{\phi}(\mathbf{z}| \mathbf{g}) = \sum_{n=1}^{N} \pi_{n}\, p\!\left(\mathbf{z}|\boldsymbol{\mu}_{\phi,n},\kappa_{\phi,n}\right),
\end{equation}
with $\pi_n \ge 0,\ \sum_{n}\pi_n=1$, where $\boldsymbol{\mu}_{\phi,n}\in\mathbb{S}^2$ and $\kappa_{\phi,n}\!\ge\!0$ are the encoder outputs for component $n$, and $\{\pi_n\}$ are mixture weights (e.g., softmax outputs of the encoder) \footnote{For $N\!=\!1$ this reduces to the single-vMF posterior used in the single-source case.}
Given $\{\mathbf{z}_n\}_{n=1}^N$ sampled via the vMF reparameterization, the physics-based decoder defines, for each microphone pair $k$, a mixture likelihood over delay bins by reusing the single-source Gaussian logits \eqref{eq:logitsoutput} and softmax normalization \eqref{eq:est_dist}, as
\begin{equation}
p(\boldsymbol{\tau}_k | \mathbf{z}) = \sum_{n=1}^{N} \pi_{n}^{\text{dec}}\; p(\boldsymbol{\tau}_k | \mathbf{z}_n),
\end{equation}
with $\pi_{n}^{\text{dec}} \ge 0,\ \sum_{n}\pi_{n}^{\text{dec}}=1$, where $\{\pi_{n}^{\text{dec}}\}$ are fixed hyperparameters (e.g., uniform) or constrained trainable scalars -- implemented with proper constraints in the loss function \cite{Luenberger2015}.

To compare the decoder likelihood against the input-induced distribution from \eqref{eq:input_dist}, a weighted physics loss with per-source activity/assignment masks $\mathbf{A}_{\mathbf{x}_n}$ can be adopted,
\begin{equation}
\mathcal{L}_{\text{physics}} = 
\mathbb{E}\!\left[
\sum_{k}
\sum_{n=1}^{N}
\pi_{n}^{\text{dec}}\,
\mathbf{A}_{\mathbf{x}_n}\;
D_{\mathrm{KL}}\!\left(
p(\boldsymbol{\tau}_k | \mathbf{g}_k)
\ || \
p(\boldsymbol{\tau}_k | \mathbf{z}_n)
\right)
\right],
\end{equation}
where $\mathbf{A}_{\mathbf{x}_n}\in[0,1]$ masks time frames (and optionally pairs) attributed to source $n$.

The KL regularizer extends the single vMF case from \eqref{eq:doaelbo} to a mixture posterior against the uniform prior on $\mathbb{S}^2$. Since a closed form is not available for a general mixture,
$
\mathcal{L}_{\mathrm{KL}} = D_{\mathrm{KL}}\!\big(q_{\phi}(\mathbf{z}|\mathbf{g}) \,\|\, \mathrm{Uniform}(\mathbb{S}^2)\big)
$
can be estimated either by Monte Carlo or by an upper bound. 

\section{Conclusions}

We proposed a solution for unsupervised sound source tracking where a variational encoder is combined with a physics decoder. We derived the theoretical foundation for both encoder and decoder, presenting an audio-tailored ELBO for the unsupervised training of the proposed system. By analyzing the ELBO hyperparameters in the limit, we show more clearly that the method is capable of modeling uncertainty and it's lower-bounded by an analogous form of the SRP method. Moreover, the architecture of the variational encoder is general, allowing for efficient parallel processing, and uses the microphones' relative positions as biases for the convolutional blocks, accelerating convergence and facilitating generalization to unknown geometries.

The proposed method is evaluated over three experiments, which differ in noise type and uncertainty of testing geometry, compared to a classical and two supervised learning-based methods. The physics-guided variational model consistently outperforms the classical approach and presents comparable performance, in terms of root mean square angular error, to its supervised counterparts. Importantly, it retains higher generalization capacity given its variational nature, outperforming one of the supervised learning approaches when trained with array geometries that differ from evaluation. The proposed approach is also more robust to uncertainty in the microphone array metadata when compared to a state-of-the-art supervised model. Additionally, we analyze the models' complexity -- showing that the proposed approach has similar complexity to that of the state-of-the-art -- and the behavior of the estimated concentration parameter, connecting its interpretation to the notion of uncertainty.

Beyond this paper's scope, we suggest the extension of the proposed model to a multi-source scenario. In Section~\ref{sec:multi}, we derive the basic theory for the multi-source operation, which can be used as a starting point for future work. Furthermore, another extension would be the generalization to number of microphones, leveraging the parallel architecture as done in \cite{grinstein2024neuralsrp}. To maintain unsupervised operation, we suggest the proposed encoder to be extended as the combination of a continuous and categorical encoder, similar to \cite{fiorio2025clustering}. The categorical part should treat the number of microphones as a random variable. The continuous encoder and physics-based decoder can inherit most of this paper's proposed model, with minor changes.

\section*{Acknowledgment}
This work was supported by the Robust AI for SafE (radar) signal processing (RAISE) collaboration framework between Eindhoven University of Technology and NXP Semiconductors, including a Privaat-Publieke Samenwerkingen-toeslag (PPS) supplement from the Dutch Ministry of Economic Affairs and Climate Policy.

The authors would like to thank Frans Widdershoven, Wim van Houtum, and Wu Yan for the discussions that enriched this work.

\bibliographystyle{IEEEtran}
\bibliography{template}

\end{document}